\documentclass[reprint, 10pt, amsmath, amssymb, floatfix, aps, prd, superscriptaddress, nofootinbib, longbibliography, nobibnotes, onecolumn]{revtex4-1}

\pdfoutput=1
\usepackage[utf8]{inputenc}
\usepackage{graphicx}
\graphicspath{{figures/}}
\usepackage{amsmath}
\usepackage{amsfonts}
\usepackage{amssymb}
\usepackage{multirow}
\usepackage{CJKutf8}
\usepackage{color}
\usepackage{acro}
\usepackage{adjustbox}
\usepackage{array}
\usepackage{natbib}
\usepackage{dblfnote}
\DFNalwaysdouble
\usepackage{slashed}
\usepackage{dcolumn}
\usepackage{geometry}
\usepackage{subfig}
\usepackage{overpic}
\usepackage{caption}
\usepackage[colorlinks=true, linkcolor=red, citecolor=blue]{hyperref}
\usepackage[capitalise]{cleveref}
\usepackage{orcidlink}

\def\({\left(}
\def\){\right)}
\def\[{\left[}
\def\]{\right]}
\def\be{\begin{eqnarray}}
\def\ee{\end{eqnarray}}

\DeclareAcronym{GW}{
  short = GW ,
  long = gravitational wave ,
  short-plural = s
}
\DeclareAcronym{LIGO}{
  short = LIGO ,
  long = Laser Interferometer Gravitational-wave Observatory ,
  short-plural =
}
\DeclareAcronym{LISA}{
  short = LISA ,
  long = Laser Interferometer Space Antenna ,
  short-plural =
}
\DeclareAcronym{SKA}{
  short = SKA ,
  long = Square Kilometre Array ,
  short-plural =
}

\DeclareAcronym{SNR}{
	short = SNR ,
	long = signal-to-noise ratio ,
	short-plural = s
}

\DeclareAcronym{PTA}{
	short = PTA ,
	long = pulsar timing array ,
	short-plural =
}

\DeclareAcronym{FLRW}{
  short = FLRW ,
  long = Friedmann--Lemaitre--Robertson--Walker ,
  short-plural =
}

\DeclareAcronym{SIGW}{
	short = SIGW ,
	long = scalar-induced gravitational wave ,
	short-plural =  s
}

\DeclareAcronym{PBH}{
	short = PBH ,
	long = primordial black hole ,
	short-plural =  s
}

\DeclareAcronym{SMBHB}{
  short = SMBHB ,
  long = supermassive black hole binary ,
  short-plural = s
}

\DeclareAcronym{KDE}{
  short = KDE ,
  long = kernel density estimator ,
  short-plural = s
}

\DeclareAcronym{BPBHM}{
  short = BPBHM ,
  long = binary primordial black hole merger,
  short-plural = s
}

\DeclareAcronym{CMB}{
	short = CMB ,
	long = cosmic microwave background ,
	short-plural =
}
\DeclareAcronym{DM}{
	short = DM ,
	long = dark matter ,
	short-plural =
}

\DeclareAcronym{BBN}{
	short = BBN ,
	long = Big-Bang nucleosynthesis ,
	short-plural =
}

\DeclareAcronym{LN}{
	short = LN ,
	long = log-normal  ,
	short-plural =
}

\DeclareAcronym{BPL}{
	short = BPL ,
	long = broken power-law ,
	short-plural =
}

\DeclareAcronym{SGWB}{
	short = SGWB ,
	long = stochastic gravitational-wave background ,
	short-plural =  s
}

\DeclareAcronym{LSS}{
	short = LSS ,
	long = large scale structure ,
	short-plural =
}

\DeclareAcronym{RD}{
	short = RD ,
	long = radiation-dominated ,
	short-plural =
}

\DeclareAcronym{PLS}{
	short = PLS ,
	long = power low sensitivity ,
	short-plural =
}

\DeclareAcronym{MAP}{
	short = MAP ,
	long = maximum a posterior ,
	short-plural =
}

\DeclareAcronym{BAO}{
	short = BAO ,
	long = baryon acoustic oscillation ,
	short-plural =
}

\begin{document}
\begin{CJK*}{UTF8}{gbsn}
\title{Probing small-scale anisotropic inflation with stochastic gravitational-wave background}

\author{Yu-Ting Kuang (况宇庭)\orcidlink{0000-0002-7431-4454}}
\affiliation{Institute of High Energy Physics, Chinese Academy of Sciences, Beijing 100049, China}
\affiliation{University of Chinese Academy of Sciences, Beijing 100049, China}

\author{Jing-Zhi Zhou (周敬之)\orcidlink{0000-0003-2792-3182}}
\affiliation{Center for Joint Quantum Studies and Department of Physics,
School of Science, Tianjin University, Tianjin 300350, China}

\author{Zhe Chang (常哲)\orcidlink{0000-0002-9720-803X}}
\affiliation{Institute of High Energy Physics, Chinese Academy of Sciences, Beijing 100049, China}
\affiliation{University of Chinese Academy of Sciences, Beijing 100049, China}

\author{Di Wu (吴迪)\orcidlink{0000-0001-7309-574X}}
\email{wudi@ucas.ac.cn}
\affiliation{School of Fundamental Physics and Mathematical Sciences, Hangzhou Institute for Advanced Study, UCAS, Hangzhou 310024, China}

\begin{abstract}
In June 2023, multiple \ac{PTA} collaborations provided evidence for the existence of a \ac{SGWB}. As a significant source of the \acp{SGWB}, \acp{SIGW} receive extensive attention. We explore the influence of anisotropic primordial power spectra on second-order \acp{SIGW} and derive explicit expressions for the energy density spectra. For specific anisotropic inflation models, we analyze the impacts of Finslerian inflation and gauge field inflation models on \acp{PTA} and the \ac{LISA} and generalize the findings to model-independent scenarios. Our results indicate that current \ac{PTA} observations cannot rule out the existence of small-scale anisotropic primordial perturbations.

\end{abstract}

\maketitle
\end{CJK*}
\acresetall

\section{Introduction}
In June 2023, the \ac{PTA} collaborations NANOGrav~\cite{NANOGrav:2023gor}, EPTA~\cite{EPTA:2023fyk}, PPTA~\cite{Reardon:2023gzh}, and CPTA~\cite{Xu:2023wog} reported positive evidence for an isotropic, stochastic background of \acp{GW} within the nanohertz frequency range. The astrophysical origin of the \ac{SGWB} is primarily attributed to \ac{SMBHB}~\cite{Shen:2023pan, Ellis:2023dgf}. Furthermore, the \ac{PTA} observation has provided a new tool for probing new physics~\cite{NANOGrav:2023hvm}. Various sources of the \ac{SGWB} have been investigated to explain the origin of current \ac{PTA} observations. The cosmological explanations include cosmic strings and domain walls~\cite{Ellis:2023tsl,Kitajima:2023cek,Bai:2023cqj}, cosmological phase transition \acp{GW}~\cite{Fujikura:2023lkn,Bringmann:2023opz}, and \acp{SIGW}~\cite{Ananda:2006af,Chang:2023vjk,Balaji:2023ehk,Cai:2023dls,Wang:2023sij,Wang:2023ost,Zhou:2021vcw}. Afzal \emph{et al}.~\cite{NANOGrav:2023hvm} has analyzed the possibility that different sources of the \ac{SGWB} dominate current PTA observations. Their results show that the \acp{SIGW} yield the highest Bayes factor, indicating that \acp{SIGW} are among the most likely sources of the \ac{SGWB} in the \ac{PTA} frequency band.

\acp{SIGW} originate from primordial curvature perturbations, with large amplitudes, on small scales. More precisely, based on large-scale cosmological observations, such as \ac{CMB} and \acp{LSS}, the amplitude of the power spectrum corresponding to primordial curvature perturbations is approximately $2\times 10^{-9}$~\cite{Planck:2018vyg}. However, on small scales ($\lesssim$1 Mpc), current cosmological observations impose significantly weaker constraints on primordial curvature perturbations~\cite{Bringmann:2011ut}. Large-amplitude primordial curvature perturbations on small scales can re-enter the horizon after inflation, thereby exciting higher-order \acp{SIGW} with significant observable effects. Since the second-order \acp{SIGW} are generated by primordial scalar perturbations for a specific inflation model, the parameter space of the inflation model can be constrained by current and future  \ac{SGWB} observations. Over the past two years, the \ac{SGWB} in the PTA frequency band, dominated by \acp{SIGW}, has been systematically studied. Relevant research includes investigations into primordial non-Gaussianity~\cite{Liu:2023ymk,Franciolini:2023pbf,Li:2024zwx}, varying sound speed~\cite{Balaji:2023ehk}, different epochs of the Universe ~\cite{Liu:2023pau,Zhu:2023gmx,Domenech:2024rks}, and corrections from third-order \acp{SIGW}~\cite{Chang:2023vjk}.

In current studies on \acp{SIGW} and PTA observations, a statistically isotropic primordial power spectrum on small scales is commonly assumed. This assumption implies that the spectrum $\mathcal{P}_{\zeta}(k)$ depends only on the magnitude of the wavenumber $\mathbf{k}$, and not on its direction. Given that present PTA data provide evidence for an isotropic \ac{SGWB}, this assumption of small-scale isotropy appears quite reasonable. However, notably, this prior assumption is not mandatory. The exploration of the primordial power spectra exhibiting anisotropy on small scales is grounded in strong physical considerations.
As indicated in previous studies~\cite{Wang:2023ost,Bartolo:2019zvb,Li:2023qua}, \acp{SIGW} originate from extremely high redshifts, corresponding to extremely small horizon scales. Due to the limited angular resolution of current detectors, the signal along any line of sight represents an ensemble average over numerous such horizon patches. Consequently, current PTA observations are incapable of resolving the anisotropy in the primordial power spectra at small scales. Constraining such anisotropic features remains an open question. In this study, we investigate the second-order \acp{SIGW} generated by small-scale ($\lesssim$1 Mpc) statistically anisotropic primordial scalar perturbations. Models capable of producing anisotropic primordial perturbations can be primarily categorized into two types: (i) models in which additional fields are introduced during inflation to generate anisotropy at the quantum scale~\cite{Ackerman:2007nb,Soda:2012zm,Maleknejad:2012fw,Dimastrogiovanni:2010sm,Chen:2022qec} and (ii) those incorporating an anisotropic spacetime background, primarily focusing on Finslerian inflation~\cite{Pfeifer:2011xi,Li:2014taa,Russell:2015gwa,Chang:2018bjg}. Here, we investigate the impact of anisotropic primordial power spectra on \acp{SIGW} using a model-independent approach, specifically a parametric method. When the primordial power spectrum exhibits anisotropy on small scales, the energy density spectra of second-order \acp{SIGW} are also anisotropic on small scales. Because the small-scale anisotropy of \acp{SIGW} cannot be observed by current PTAs, we derive the isotropic energy density spectrum of \acp{SIGW} by spatially averaging the anisotropic \acp{SIGW} on small scales. Evidently, the spatial averaging does not eliminate the influence of the anisotropic parameters. The potential small-scale anisotropic primordial power spectra can be constrained by current \acp{PTA} observations of the isotropic energy density spectrum of \acp{SIGW}.

The remainder of this paper is organized as follows. In Sec.~\ref{sec:2.0}, we review the calculations of the second-order \acp{SIGW}. In Sec.~\ref{sec:3.0}, we calculate the second-order \acp{SIGW} for cases of anisotropic primordial power spectra. In Sec.~\ref{sec:4.0}, we present several models of anisotropic primordial power spectra and use \ac{PTA}+\ac{CMB}+\ac{BAO} data to constrain the anisotropy parameters. We summarize our results in Sec.~\ref{sec:5.0}.

\section{Review of SIGW}\label{sec:2.0}
When studying \acp{SIGW}, the effects of primordial vector and tensor perturbations are typically ignored. Primordial vector perturbations, in particular, decay as $a^{-2}$, thus showing contribution~\cite{Bassett:2005xm}. While primordial tensor perturbations are well-constrained on large scales, their amplitude can be enhanced at small scales using specific inflationary models~\cite{Cai:2020ovp,Gorji:2023ziy,Oikonomou:2023qfz}. In such scenarios, primordial tensor perturbations can significantly impact the induced \acp{GW}~\cite{Chang:2022vlv}. In this paper, we focus solely on the large primordial scalar perturbations on small scales. The perturbed metric in the \ac{FLRW} spacetime with Newtonian gauge can be written as
\begin{equation}
    \begin{aligned}
    \label{eq:FLRW}
        \mathrm{d}s^{2}=&a^{2}\left(-\left(1+2 \phi^{(1)}\right) \mathrm{d} \eta^{2}+\left(\left(1-2 \psi^{(1)}\right) \delta_{i j}+\frac{1}{2} h_{i j}^{(2)}\right)\mathrm{d} x^{i} \mathrm{d} x^{j}\right) \ ,
    \end{aligned}
\end{equation}
where $\phi^{(1)}$ and $\psi^{(1)}$ denote first-order scalar perturbations, and $h_{ij}^{(2)}$ is the second-order tensor perturbation. In this study, we investigated \acp{SIGW} generated during the \ac{RD} era ($\omega = 1/3$ and $c_s = 1/\sqrt{3}$). The equations of motion of first-order scalar perturbations are given by
\begin{equation}
    \begin{aligned}
    \label{eq:1-order perturbations}
        &6\psi^{(1)\prime\prime}(\mathbf{x},\eta) + \Delta(3\phi^{(1)}(\mathbf{x},\eta) - 5\psi^{(1)}(\mathbf{x},\eta))+ 6\mathcal{H}(\phi^{(1)\prime}(\mathbf{x},\eta) + 3\psi^{(1)\prime}(\mathbf{x},\eta)) = 0\ ,\\
        &\psi^{(1)}(\mathbf{x},\eta) - \phi^{(1)}(\mathbf{x},\eta) = 0 \ ,
    \end{aligned}
\end{equation}
where the prime stands for derivative with respect to the conformal time $\eta$, and $\mathcal{H} = aH = 1/\eta$ is the conformal Hubble parameter. In momentum space, the solutions of Eq.~(\ref{eq:1-order perturbations}) can be written as~\cite{Inomata:2020cck}
\begin{equation}\label{eq:T1}
    \begin{aligned}
        &\psi^{(1)}(\mathbf{k},\eta) = \phi^{(1)}(\mathbf{k},\eta) = T_{\phi}(k\eta)\phi_{\mathbf{k}}= \frac{9}{(k\eta)^2}\left(\frac{\sin(k\eta/\sqrt{3})}{k\eta/\sqrt{3}} - \cos(k\eta/\sqrt{3})\right)\frac{2}{3}\zeta_{\mathbf{k}}\ ,
    \end{aligned}
\end{equation}
where $T_{\phi}(x)$ is the transfer function of $\phi^{(1)}$, and $\zeta_{\mathbf{k}}$ represents the primordial curvature perturbation. We set $k=|\mathbf{k}|$ in Eq.~(\ref{eq:T1}).

By expanding the Einstein field equations to the second order, we obtain the equation of motion of second-order \acp{SIGW}:
\begin{equation}
    \begin{aligned}
    \label{eq:h of x}
        h_{ij}^{(2)\prime\prime}(\mathbf{x},\eta) + 2\mathcal{H}h_{ij}^{(2)\prime}(\mathbf{x},\eta) - \Delta h_{ij}^{(2)}(\mathbf{x},\eta)
        = -4S^{(2)}_{ij}(\mathbf{x},\eta)\ ,
    \end{aligned}
\end{equation}
where the source term $S^{(2)}_{ij}(\mathbf{x},\eta)$ is given by
\begin{equation}\label{eq:S2h}
    \begin{aligned}
        S^{(2)}_{ij}(\mathbf{x},\eta)=  \Lambda^{rs}_{ij}\Big( 3\phi^{(1)}\partial_r\partial_s\phi^{(1)} + \frac{2}{\mathcal{H}}\phi^{(1)\prime}\partial_r\partial_s\phi^{(1)} + \frac{1}{\mathcal{H}^2}\phi^{(1)\prime}\partial_r\partial_s\phi^{(1)\prime}\Big)\ .
    \end{aligned}
\end{equation}
We have simplified the above equation using the relation $\Lambda^{rs}_{ij}\partial_r\phi\partial_s\phi = - \Lambda^{rs}_{ij}\phi\partial_r\partial_s\phi$, where
\begin{equation}
    \begin{aligned}
        \Lambda^{rs}_{ij} = \mathcal{T}^{r}_{i}\mathcal{T}^{s}_{j} - \frac{1}{2}\mathcal{T}_{ij}\mathcal{T}^{rs}
    \end{aligned}
\end{equation}
is the transverse and traceless operator, and $\mathcal{T}_{ij} = \delta_{ij} - \partial_i\Delta^{-1}\partial_j$ is the traceless operator.
The Fourier transform of Eq.~(\ref{eq:h of x}) yields
\begin{equation}
    \begin{aligned}
    \label{eq:h of p}
        h^{\lambda,(2)\prime\prime}_{\mathbf{k}}(\eta) + 2\mathcal{H}h^{\lambda,(2)\prime}_{\mathbf{k}}(\eta) + k^2 h^{\lambda,(2)}_{\mathbf{k}}(\eta)
        =-4S^{\lambda,(2)}_{\mathbf{k}}(\eta)\ ,
    \end{aligned}
\end{equation}
where the source term $S^{\lambda,(2)}_{\mathbf{k}}(\eta)$ is given by
\begin{equation}
    \begin{aligned}
    \label{eq:source in p}
        S^{\lambda,(2)}_{\mathbf{k}}(\eta) = - \int \frac{\mathrm{d} ^3\mathbf{p}}{(2\pi)^{3/2}}e^{\lambda,ij}(\mathbf{k})p_ip_j\Big(2\phi_\mathbf{p}\phi_{\mathbf{k - p}}+ (\mathcal{H}^{-1}\phi^{\prime}_\mathbf{p} + \phi_{\mathbf{p}})(\mathcal{H}^{-1}\phi^{\prime}_{\mathbf{k-p}} + \phi_{\mathbf{k - p}})\Big)\ .
    \end{aligned}
\end{equation}
In Eq.~(\ref{eq:source in p}), $e^{\lambda,ij}(\mathbf{k})$ represents the polarization tensor. The energy density spectrum of second-order \acp{SIGW} can be calculated using the following formula:
\begin{equation}
    \begin{aligned}
    \label{eq:omegaGW}
        \Omega_{\mathrm{GW}}(k,\eta) = \frac{\rho_{\mathrm{GW}}(k,\eta)}{\rho_{\mathrm{tot}}(\eta)} = \frac{1}{24}\left(\frac{k}{\mathcal{H}}\right) ^2\mathcal{P}_h(k,\eta)\ ,
    \end{aligned}
\end{equation}
where $\mathcal{P}_h(k,\eta)$ represents the power spectrum of  second-order \acp{SIGW} and is defined as
\begin{equation}
    \begin{aligned}
    \label{eq:Ph}
        \left<h^{\lambda,(2)}_{\mathbf{k}}(\eta)h^{\lambda^\prime,(2)}_{\mathbf{k}^\prime}(\eta)\right> = \delta^{\lambda\lambda^\prime}\delta^3(\mathbf{k + k^\prime})\frac{2\pi^2}{k^3}\mathcal{P}^{(2)}_h(k,\eta)\ .
    \end{aligned}
\end{equation}
As shown by Kohri and Terada~\cite{Kohri:2018awv}, the power spectrum of second-order \acp{SIGW} $\mathcal{P}^{(2)}_h(k,\eta)$ is given by
\begin{equation}
    \begin{aligned}
    \label{eq:analise function pf Ph}
        \mathcal{P}^{(2)}_h(k,\eta) = 4\int^\infty_0 \mathrm{d} v\int^{1+v}_{|1-v|} \mathrm{d} u\left(\frac{4v^2 - (1 + v^2 - u^2)^2}{4uv}\right)^2 \left(I(v,u,x)\right)^2\mathcal{P}_\zeta(kv)\mathcal{P}_\zeta(ku)\ ,
    \end{aligned}
\end{equation}
where $\mathcal{P}_{\zeta}(k)$ is the power spectrum of primordial curvature perturbations. We define $x = k\eta$, $u = |\mathbf{k-p}|/k$ and $v = p/k$. The kernel function $I(v,u,x)$ is expressed as
\begin{equation}\label{eq:I2}
    \begin{aligned}
        I\left( u,v,x \right)=\frac{4}{k^2} \int_{0}^{x} \mathrm{d} \bar{x} \left( \frac{\bar{x}}{x}\sin\left( x-\bar{x} \right) f\left( u,v,\bar{x} \right)  \right)  \ ,
    \end{aligned}
\end{equation}
where the function $f(v,u,x)$, derived from the source term in Eq.~(\ref{eq:source in p}), can be expressed as
\begin{equation}
    \begin{aligned}
        f(v,u,x) &= \frac{12}{u^3v^3x^6}\left(18uvx^2\cos\frac{ux}{\sqrt{3}}\cos\frac{vx}{\sqrt{3}} + (54 - 6(u^2 + v^2)x^2 + u^2v^2x^4)\sin\frac{ux}{\sqrt{3}}\sin\frac{vx}{\sqrt{3}} \right. \\
        & \left. + 2\sqrt{3}ux(v^2x^2 - 9)\cos\frac{ux}{\sqrt{3}}\sin\frac{vx}{\sqrt{3}}+ 2\sqrt{3}vu(u^2x^2 - 9)\sin\frac{ux}{\sqrt{3}}\cos\frac{vx}{\sqrt{3}}\right)\ .
    \end{aligned}
\end{equation}
Evaluation of the integral in Eq.~(\ref{eq:I2}) yields the analytical expression for the kernel function $I(u,v,x)$~\cite{Kohri:2018awv}:
\begin{eqnarray}
        k^2 I\left(u,v,x\to \infty \right)& =&
        \frac{27 (u^2 + v^2 - 3)}{u^3 v^3 x}
        \bigg(
            \sin x
            \big(
                -4 u v + (u^2 + v^2 - 3)
                \ln \left| \frac{3 - (u + v)^2}{3 - (u - v)^2} \right|
            \big) \nonumber\\
        &-& \pi (u^2 + v^2 - 3) \Theta(v + u - \sqrt{3}) \cos x
        \bigg) \ . \label{eq:11I}
\end{eqnarray}
In Eq.~(\ref{eq:11I}), the approximations: $\lim_{x\to\pm \infty} \mathrm{Si}(x)=\pm \pi/2$ and $\lim_{x\to \infty} \mathrm{Ci}(x)=0$ are used. Using the analytical expression of the kernel function given in Eq.~(\ref{eq:11I}), together with the results for the corresponding power and energy density spectra provided in Eq.~(\ref{eq:analise function pf Ph}) and Eq.~(\ref{eq:omegaGW}), respectively, we obtain the formula for the energy density spectrum of second-order \acp{SIGW} during the \ac{RD} era:
\begin{align}
    \Omega_{\mathrm{GW}}(k) &= \int_{0}^{\infty} \mathrm{d}v \int_{|1-v|}^{1+v} \mathrm{d}u\,
    \mathcal{P}_{\zeta}(uk) \mathcal{P}_{\zeta}(vk) \notag \\
    &\times  \frac{3}{1024 u^8 v^8} (u^2 + v^2 - 3)^2 \left[4v^2 - (1 + v^2 - u^2)^2\right]^2 \notag \\
    &\times \left\{ \left[(u^2 + v^2 - 3) \ln \left| \frac{3 - (u + v)^2}{3 - (u - v)^2} \right| - 4uv \right]^2  +   \pi^2 (u^2 + v^2 - 3)^2 \Theta \Big(u + v - \sqrt{3} \Big) \right\} \ ,
    \label{eq:1Oss}
\end{align}
where the squared kernel function is subjected to oscillatory averaging using the relations $\sin^2 x\sim 1/2$ and $\cos^2 x\sim 1/2$~\cite{Yuan:2021qgz}. Given a specific form of the power spectrum of primordial curvature perturbations $\mathcal{P}_{\zeta}(k)$, we can use Eq.~(\ref{eq:1Oss}) to calculate the energy density spectrum of second-order \acp{SIGW} during the \ac{RD} era.

\section{Anisotropic primordial power spectrum and \acp{SIGW}}\label{sec:3.0}

\subsection{ Energy density spectrum}\label{sec:3.1}
One of the primary methods of inducing anisotropy in primordial curvature perturbations is to introduce an anisotropic vector field during the inflationary period. Upon coupling between the vector field with the inflaton field~\cite{Maleknejad:2012as,Ackerman:2007nb,Soda:2012zm,Maleknejad:2012fw,Dimastrogiovanni:2010sm,Chen:2022qec}, the power spectrum of primordial curvature perturbations becomes anisotropic. Another key approach is to modify the background spacetime during inflation to the Finsler spacetime~\cite{Pfeifer:2011xi,Li:2014taa,Russell:2015gwa,Chang:2018bjg}; through calculations similar to those in performed traditional inflation models, this second approach results in anisotropic primordial curvature perturbations.

In this study, we adopt a model-independent approach by analyzing the anisotropic primordial power spectrum through parameterization. The small-scale anisotropic primordial power spectrum is expressed as
\begin{equation}
    \begin{aligned}
    \label{eq:anisotropic spectrum}  \mathcal{P}^{\hat{\mathbf{n}}}_\zeta(\mathbf{k}) = \mathcal{P}_{0,\zeta}(k)\sum^{\infty}_{l=0}(-i)^l(2l + 1)A_l(k)\mathcal{P}_l(\hat{\mathbf{n}}\cdot\hat{\mathbf{k}})\ ,
    \end{aligned}
\end{equation}
where $\hat{\mathbf{n}}$ is the direction of anisotropy, $\hat{\mathbf{k}}$ is the unit vector along $\mathbf{k}$, and $\mathcal{P}_l$ is the Legendre polynomial of order $l$. When $A_0(k) = 1$ and $A_l(k) = 0$ for $l\geq 1$, the primordial power spectrum reduces to $\mathcal{P}_{0,\zeta}(k)$, corresponding to the isotropic case. In a previous study, cases with non-zero $A_0$ and $A_2$ have been investigated~\cite{Chen:2022qec}. In the present study, we consider the anisotropic primordial power spectrum with $A_l\neq 0$ for $l \leq 4$. In this case, Eq.~(\ref{eq:anisotropic spectrum}) can be rewritten as
\begin{equation}
    \begin{aligned}
    \label{eq:anisotropic spectrum with C}
        \mathcal{P}^{\hat{\mathbf{n}}}_\zeta(\mathbf{k}) = \mathcal{P}_{0,\zeta}(k)\sum^{4}_{l=0}C_l(k)\mathcal{P}_l(\hat{\mathbf{n}}\cdot\hat{\mathbf{k}})\ ,
    \end{aligned}
\end{equation}
where $C_l = (-i)^l(2l + 1)A_l(k)$. For simplicity, in the following discussion, we neglect the dependence of $C_l$ on $k$, assuming that $C_l$ is constant on small scales.

Because the current angular resolution of \ac{GW} observational data is insufficient for detecting small-scale anisotropies, spatial averaging of the anisotropic power spectrum must be performed to obtain an isotropic energy density spectrum. Following spatial averaging, the power spectrum of the second-order \acp{GW}  induced by anisotropic primordial scalar perturbations  can be expressed as
\begin{equation}
    \begin{aligned}
    \label{eq:analise function pf Ph in anisotropic}
        \mathcal{P}^{(2)}_h(k,\eta) =& 4\int^{2\pi}_0 \mathrm{d} \phi_n\int^{\pi}_0\frac{\sin(\theta_n)}{4\pi} \mathrm{d} \theta_n\int^{2\pi}_0\frac{\mathrm{d}\phi_p}{2\pi}\int^\infty_0 \mathrm{d} v \int^{1+v}_{|1-v|} \mathrm{d} u\left(\mathcal{P}_{0,\zeta}(kv)\sum_{l_1=0}^4 C_{l_1}\mathcal{P}_{l_1}(\hat{\mathbf{n}}\cdot \hat{\mathbf{p}})\right)\\
        &\left(\mathcal{P}_{0,\zeta}(ku)\sum_{l_2=0}^4 C_{l_2}\mathcal{P}_{l_2}(\hat{\mathbf{n}}\cdot \widehat{\mathbf{k - p}})\right)\left(\frac{4v^2 - (1 + v^2 - u^2)^2}{4uv}\right)^2I^2(v,u,x)\ ,
    \end{aligned}
\end{equation}
where $\phi_n$ and $\theta_n$ are the azimuth and elevation of $\hat{\mathbf{n}}$, and $\phi_p$ is the azimuth of $\hat{\mathbf{p}}$.
For convenience, we introduce
\begin{equation}
    \begin{aligned}
    \label{eq:Ql1l2}
    Q_{l_1,l_2} = &\int^{2\pi}_0 \mathrm{d} \phi_n\int^{\pi}_0\sin(\theta_n) \mathrm{d} \theta_n\mathcal{P}_{l_1}(\hat{\mathbf{n}}\cdot\hat{\mathbf{p}})\mathcal{P}_{l_2}(\hat{\mathbf{n}}\cdot\widehat{\mathbf{k-p}})\ ,
    \end{aligned}
\end{equation}
which allows Eq.~(\ref{eq:analise function pf Ph in anisotropic}) to be rewritten as
\begin{equation}
    \begin{aligned}
    \label{eq:simplified analise function pf Ph in anisotropic}
        \mathcal{P}^{(2)}_h(k,\eta) =& 4\int^\infty_0 \mathrm{d} v\int^{1+v}_{|1-v|} \mathrm{d} u\left(\frac{4v^2 - (1 + v^2 - u^2)^2}{4uv}\right)^2 \left(\sum\limits_{\hspace{10pt} l_1,l_2=0}^4 C_{l_1}C_{l_2}Q_{l_1,l_2}\right)\\
        &\times \left(I(v,u,x)\right)^2\mathcal{P}_{0,\zeta}(kv)\mathcal{P}_{0,\zeta}(ku)\ .
    \end{aligned}
\end{equation}
A comparingson of Eq.~(\ref{eq:analise function pf Ph}) with Eq.~(\ref{eq:simplified analise function pf Ph in anisotropic}) shows that the impact of the anisotropic power spectrum is only reflected in the additional terms in the first line of Eq.~(\ref{eq:simplified analise function pf Ph in anisotropic}). The analytical result of $Q_{l_1,l_2}$ in Eq.~(\ref{eq:Ql1l2}) and Eq.~(\ref{eq:simplified analise function pf Ph in anisotropic}) is given by
\begin{equation}
    \begin{aligned}
    \label{eq:analytic equation of Ql1l2}
    &Q_{0,0} =  4\pi\ ,\\
    &Q_{1,1} = -2\pi\frac{u^2 + v^2 -1}{3uv}\ ,\\
    &Q_{2,2} = \pi \frac{3u^2 + 2u^2(v^2 - 3) + 3(v^2 - 1)^2}{10 u^2 v^2}\ ,\\
    &Q_{3,3} = -\pi(u^2 + v^2 -1)
    \frac{5(u^2-1)^2+5v^4 - 2v^2(u^2 + 5)}{28u^3v^3}\\
    &Q_{4,4} = \pi\Bigg(35u^8 + 20u^4(v^2 - 7)  + 20u^2(v^2 - 1)^2(v^2 - 7) \\
    &\ \ \ \ \ \ \ \ \ + 35(v^2 - 1)^4  + 6u^4(3v^4 - 30v^2 + 35) \bigg)\frac{1}{288u^4v^4}\ ,\\
    &Q_{l_1, l_2} = 0 \ \ \ (l_1 \neq l_2)\ .
    \end{aligned}
\end{equation}
Substituting Eq.~(\ref{eq:simplified analise function pf Ph in anisotropic}) into Eq.~(\ref{eq:omegaGW}) yields the corresponding expression for the energy density spectrum during the \ac{RD} era:
\begin{align}
    \Omega_{\mathrm{GW}}(k) &= \int_{0}^{\infty} \mathrm{d}v \int_{|1-v|}^{1+v} \mathrm{d}u\,
    \mathcal{P}_{\zeta}(uk) \mathcal{P}_{\zeta}(vk) \left(\sum\limits_{\hspace{10pt} l_1,l_2=0}^4 C_{l_1}C_{l_2}Q_{l_1,l_2}\right)\notag \\
    &\times  \frac{3}{1024 u^8 v^8} (u^2 + v^2 - 3)^2 \left[4v^2 - (1 + v^2 - u^2)^2\right]^2 \notag \\
    &\times \left\{ \left[(u^2 + v^2 - 3) \ln \left| \frac{3 - (u + v)^2}{3 - (u - v)^2} \right| - 4uv \right]^2  +   \pi^2 (u^2 + v^2 - 3)^2 \Theta \Big(u + v - \sqrt{3} \Big) \right\} \ .
    \label{eq:1Ossan}
\end{align}
When considering the small-scale anisotropic primordial power spectrum for $l \leq 4$, the corresponding energy density spectrum of second-order \acp{SIGW} can be directly computed using Eq.~(\ref{eq:1Ossan}). As previously discussed, current observations of the \ac{SGWB} are unable to resolve the anisotropy generated by the small-scale anisotropic primordial power spectrum. As shown in Eq.~(\ref{eq:1Ossan}), even after taking the spatial average of the energy density spectrum of \acp{SIGW}, the anisotropy parameters $C_l$ still affect the energy density spectrum. Therefore, the observations of the current \ac{SGWB} can be used to constrain the small-scale anisotropic primordial power spectrum.

\begin{figure}[htbp]
\includegraphics[width=0.8\linewidth]{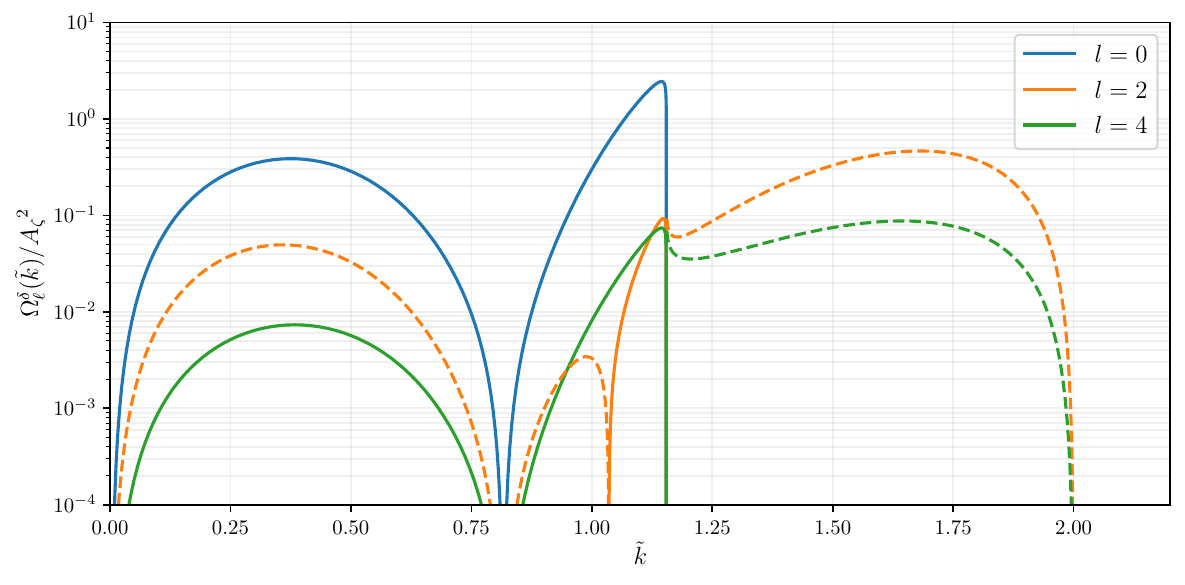}
\caption{$\Omega_l^\delta$ as a function of $\tilde{k}$, assuming $A_2=0.2$. The dashed curves denote the absolute value of the negative ratios of $l = 0, 2$.} \label{fig:omega_l}
\end{figure}

{
\subsection{Anisotropy of \acp{SIGW}}\label{sec:3.2}
In the previous subsection, we derived the explicit expression for the energy density spectrum of second-order \acp{SIGW}, which originates from the small-scale anisotropic primordial power spectrum, after performing spatial averaging. As we mentioned earlier, due to the limited precision of current observational data, such small-scale anisotropies cannot be directly detected. Nevertheless, analyzing the anisotropy of the energy density spectrum of \acp{SIGW} at small scales remains meaningful, as it may provide a window for probing small-scale anisotropic inflationary models in the future. In this subsection, we compute the anisotropy of second-order \acp{SIGW} at small scales. For simplicity, we focus on the primordial power spectrum with $l=0$ and $l=2$. Such an anisotropic primordial spectrum is typically associated with anisotropic inflationary models involving vector or gauge fields, which are discussed in Sec.~\ref{sec:4.0}. Based on Eq.~(\ref{eq:anisotropic spectrum}), the anisotropic primordial power spectrum in this scenario can be expressed as
\begin{equation}\label{eq:12p}
\mathcal{P}^{\hat{\mathbf{n}}}_\zeta(\mathbf{k}) = \mathcal{P}_{0,\zeta}(k) \left(1-5A_2(k) \right) \mathcal{P}_l(\hat{\mathbf{n}}\cdot\hat{\mathbf{k}})\ .
\end{equation}
By substituting Eq.~(\ref{eq:12p}) into Eq.~(\ref{eq:analise function pf Ph}) and simplifying, we obtain~\cite{Chen:2022qec}
\begin{equation}\label{eq:Omegan}
    \begin{aligned}
        \Omega^{\mathbf{n}}_{\text{GW}}(\eta, \mathbf{k})
        &= \sum_{l=0}^{\infty} (-i)^{l} (2l + 1)
        \Omega_{l}(\eta, k) P_{l}(\hat{\mathbf{n}} \cdot \hat{\mathbf{k}}) \ .
    \end{aligned}
\end{equation}
In Eq.~(\ref{eq:Omegan}), $\Omega_{l}(\eta, k)$ represents the multipole moment associated with second-order \acp{SIGW}, and its explicit form can be expressed as
\begin{equation}\label{eq:Oll}
    \begin{aligned}
        \Omega_{l}(x, k)
        &=\frac{x^2}{6} \int_0^\infty \mathrm{d} v \int_{|1 - v|}^{1 + v} \mathrm{d} u
        \left( \frac{4v^2 - (1 + v^2 - u^2)^2}{4uv} \right)^2
        I^2(u,v,x) \mathcal{P}_\zeta(uk) \mathcal{P}_\zeta(vk)
        \left[ \delta_{l 0} + H_l(u,v,x) \right] \ ,
    \end{aligned}
\end{equation}
where $\delta_{l 0}$ is the Kronecker delta. $H_l$ in Eq.~(\ref{eq:Oll}) are given by
\begin{eqnarray}
        H_{0}&=& A_2(uk)A_2(vk)\frac{160}{81u^2v^2}
         \left( 3u^4 + 2u^2(v^2 - 3) + 3(v^2 - 1)^2 \right) \ , \\
         H_{2}&=&\frac{32A_2(uk)}{81u^2v^2}
        \left( 3u^4 v^2 + 2u^2 v^2 (1 - 3v^2) + 3v^2(v^2 - 1)^2 \right)+\frac{32A_2(vk)}{81u^2v^2}
         \left( 3u^2(1 + v^2 - u^2)^2 - 4u^2 v^2 \right) \nonumber\\
         &-&\frac{160A_2(uk)A_2(vk)}{567}
        \frac{1}{u^2 v^2} \left( 3u^6 - 3u^4(v^2 + 1) + 3(v^2 - 1)^2(v^2 + 1)
        - u^2(3 + 2v^2 + 3v^4) \right) \ , \\
        H_{4}&=&\frac{20A_2(uk)A_2(vk)}{567u^2v^2}
        \Big( 35u^8 - 20u^6(3 + 7v^2) + 6u^4(3 + 10v^2 + 35v^4) \nonumber \\
        &+&4u^2(1 + 3v^2 + 15v^4 - 35v^6) + (v^2 - 1)^2(3 + 10v^2 + 35v^4) \Big) \ .
\end{eqnarray}
To better illustrate the small-scale anisotropy of second-order \acp{SIGW}, we consider the following form of the monochromatic primordial power spectrum:
\begin{eqnarray}\label{eq:m11}
    \mathcal{P}_{0,\zeta}(k)=A_{\zeta}k_*\delta(k-k_*) \ .
\end{eqnarray}
In this case, the multipole moment of the energy density spectrum of second-order \acp{SIGW} in Eq.~(\ref{eq:Oll}) is given by
\begin{equation}
    \begin{aligned}
        \Omega_l^\delta(\tilde{k})
        &= \Omega_{\text{iso}}^\delta(\tilde{k})
        \left( \delta_{l 0} + H_l^\delta(\tilde{k} ) \right) \ ,
    \end{aligned}
\end{equation}
where
\begin{eqnarray}
        \Omega_{\text{iso}}^\delta(\tilde{k})
        &=& \frac{3A_\zeta^2}{64}
        \left( \frac{\tilde{k}^2 - 4}{4} \right)^2
        \tilde{k}^2 (3\tilde{k}^2 - 2)^2 \Bigg(
        \pi^2 (3\tilde{k}^2 - 2)^2 \, \Theta\left( \frac{2}{\sqrt{3}} - \tilde{k} \right) \nonumber \\
        &+& \left( 4 + (3\tilde{k}^2 - 2) \ln \left| 1 - \frac{4}{3\tilde{k}^2} \right|^2 \right)
        \Theta(2 - \tilde{k}) \Bigg) \ , \\
        H_0^\delta(\tilde{k} )&=&\frac{5A_2^2}{8} \left( 8 - 12\tilde{k}^2 + 3\tilde{k}^4 \right) \ , \\
        H_2^\delta(\tilde{k} )&=&\frac{A_2}{4}  (3\tilde{k}^2 - 4)
        + \frac{5A_2^2}{56}  (8 + 6\tilde{k}^2 - 3\tilde{k}^4) \ , \\
        H_4^\delta(\tilde{k} )&=&\frac{5A_2^2}{448}  (48 + 8\tilde{k}^2 + 3\tilde{k}^4) \ .
\end{eqnarray}
Fig.~\ref{fig:omega_l} presents the results for the multipole moments of the energy density spectrum of second-order \acp{SIGW}, obtained assuming a monochromatic primordial power spectrum in Eq.~(\ref{eq:12p}) and Eq.~(\ref{eq:m11}).

}

\section{Detection of anisotropic primordial power spectrum}\label{sec:4.0}
To calculate the power spectrum of the second-order \acp{SIGW}, the specific expression of the primordial power spectrum is required. In this study,  we consider the log-normal primordial power spectrum~\cite{Zhou:2020kkf,Addazi:2022ukh,Peng:2021zon,Chen:2024gqn}:
\begin{equation}
    \begin{aligned}
        \mathcal{P}_{0,\zeta}(k) = \frac{A_\zeta}{\sqrt{2\pi\sigma^2}}\exp{\left(-\frac{\ln^2(k/k_*)}{2\sigma^2}\right)}\ ,
    \end{aligned}
\end{equation}
where $A_\zeta$ is the amplitude of the primordial power spectrum, $k_* = 2\pi f_*$ is the wavenumber at which the power spectrum exhibits a log-normal peak, and $\sigma$ characterizes the width of the log-normal distribution. The anisotropy of the primordial power spectrum on small scales is characterized by $C_l$ $(l=1,2,3,4)$ in Eq.~(\ref{eq:anisotropic spectrum with C}). The energy density of second-order \acp{SIGW} during the \ac{RD} era can be calculated using Eq.~(\ref{eq:1Ossan}). Considering the thermal history of the Universe, we obtain the current total energy density spectrum of \acp{SIGW} as~\cite{Wang:2019kaf}
\begin{equation}
    \begin{aligned}
        \Omega_{0,\mathrm{GW}}(k) = \Omega_{rad,0}\left(\frac{g_{*,\rho,e}}{g_{*,\rho,0}}\right)\left(\frac{g_{*,s,0}}{g_{*,s,e}}\right)^{4/3}\Omega_{\mathrm{GW}}(k, \eta)\ ,
    \end{aligned}
\end{equation}
where $\Omega_{rad,0} = 4.2 \times 10^{-5} h^{-2}$ is the energy density fraction of radiation today, and the dimensionless Hubble constant is $h = 0.6736$~\cite{Planck:2018vyg}; $g_{*,\rho}$ and $g_{*,s}$ are the effective numbers of relativistic degrees of freedom~\cite{Saikawa:2018rcs}. Fig.~\ref{fig:tunecl} illustrates the impact of different $C_l$ $(l=1,2,3,4)$ on the energy density spectrum of second-order \acp{SIGW} under a log-normal primordial power spectrum. In this study, we consider the following two types of anisotropic inflation models:

\begin{figure}[htbp]
\includegraphics[width=0.75\linewidth]{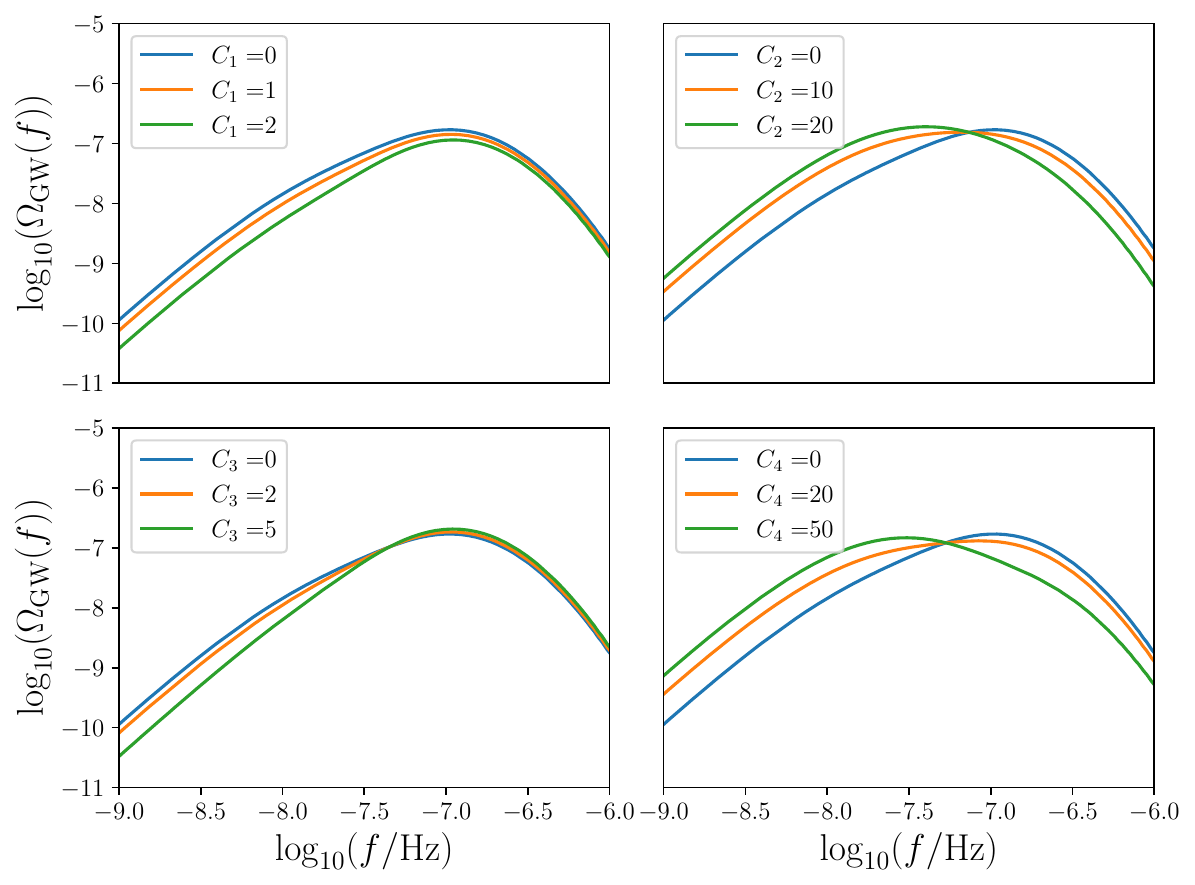}
\caption{The impact of different $C_l$ parameters on the energy density spectrum is shown. Here, $f_*$, $A_\zeta$, and $\sigma$ are fixed at $10^{-7}$ Hz, $0.5$, and $1$, respectively. In each subplot, only $C_1$, $C_2$, $C_3$, or $C_4$ is non-zero, from left to right. Different colors in the figure represent different values of the $C_l$ parameter, as indicated in the legend.} \label{fig:tunecl}
\end{figure}

\noindent
\textbf{1, Gauge field:}  Introducing a vector (gauge) field during inflation can serve as a mechanism for generating cosmic anisotropy. In this type of model, the primordial power spectrum can be parameterized as follows~\cite{Maleknejad:2012fw}:
\begin{equation}
    \begin{aligned}   \mathcal{P}^{\hat{\mathbf{n}}}_\zeta(\mathbf{k}) = \mathcal{P}_{0,\zeta}(k)\left(1 +C_2\mathcal{P}_2(\hat{\mathbf{n}}\cdot\hat{\mathbf{k}})\right) \  .
    \end{aligned}
\end{equation}

\noindent
\textbf{2, Finslerian inflation:} Modifying the spacetime background during inflation can introduce anisotropy. One of the most prominent spacetime backgrounds for this purpose is the Finsler spacetime background. In this model, the form of the anisotropic primordial power spectrum is given by~\cite{Li:2015sja}
\begin{equation}
    \begin{aligned}
        \mathcal{P}^{\hat{\mathbf{n}}}_\zeta(\mathbf{k}) = \mathcal{P}_{0,\zeta}(k)\left(1 +C_1\mathcal{P}_1(\hat{\mathbf{n}}\cdot\hat{\mathbf{k}})\right) \  .
    \end{aligned}
\end{equation}
The aforementioned two inflation models provide anisotropic primordial power spectra under the conditions $C_2\ne 0$ and $C_1\ne 0$, respectively. In the remainder of this section, we will discuss the impact of these two models on \ac{PTA} and \ac{LISA} observations separately and consider model-independent scenarios, where both $C_1$ and $C_2$ are non-zero.

\subsection{PTA observations and large-scale cosmological constraints }\label{sec:4.1}
\begin{figure}[htbp]
\includegraphics[width=0.55\linewidth]{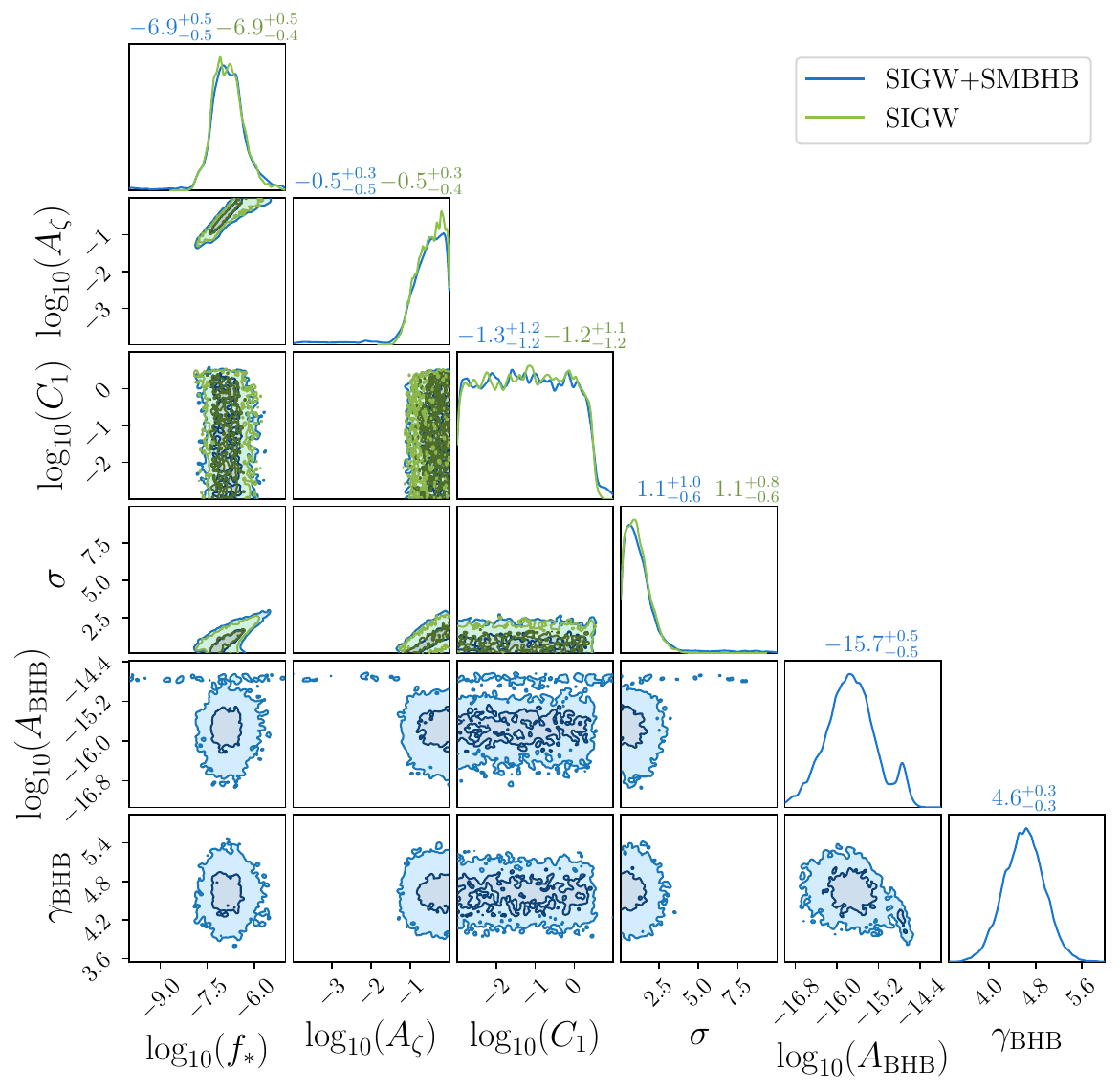}
\caption{Corner plot showing the posterior distributions with $C_1 \neq 0$. The blue and green curves correspond to the NANOGrav 15-year dataset. Off-diagonal panels display $1$-$\sigma$ and $2$-$\sigma$ confidence intervals for joint distributions, and diagonal panels provide marginal distributions with the median values and $1$-$\sigma$ uncertainties noted above each histogram.} \label{fig:corner_cl1}
\end{figure}

\begin{figure}[htbp]
\includegraphics[width=0.5\linewidth]{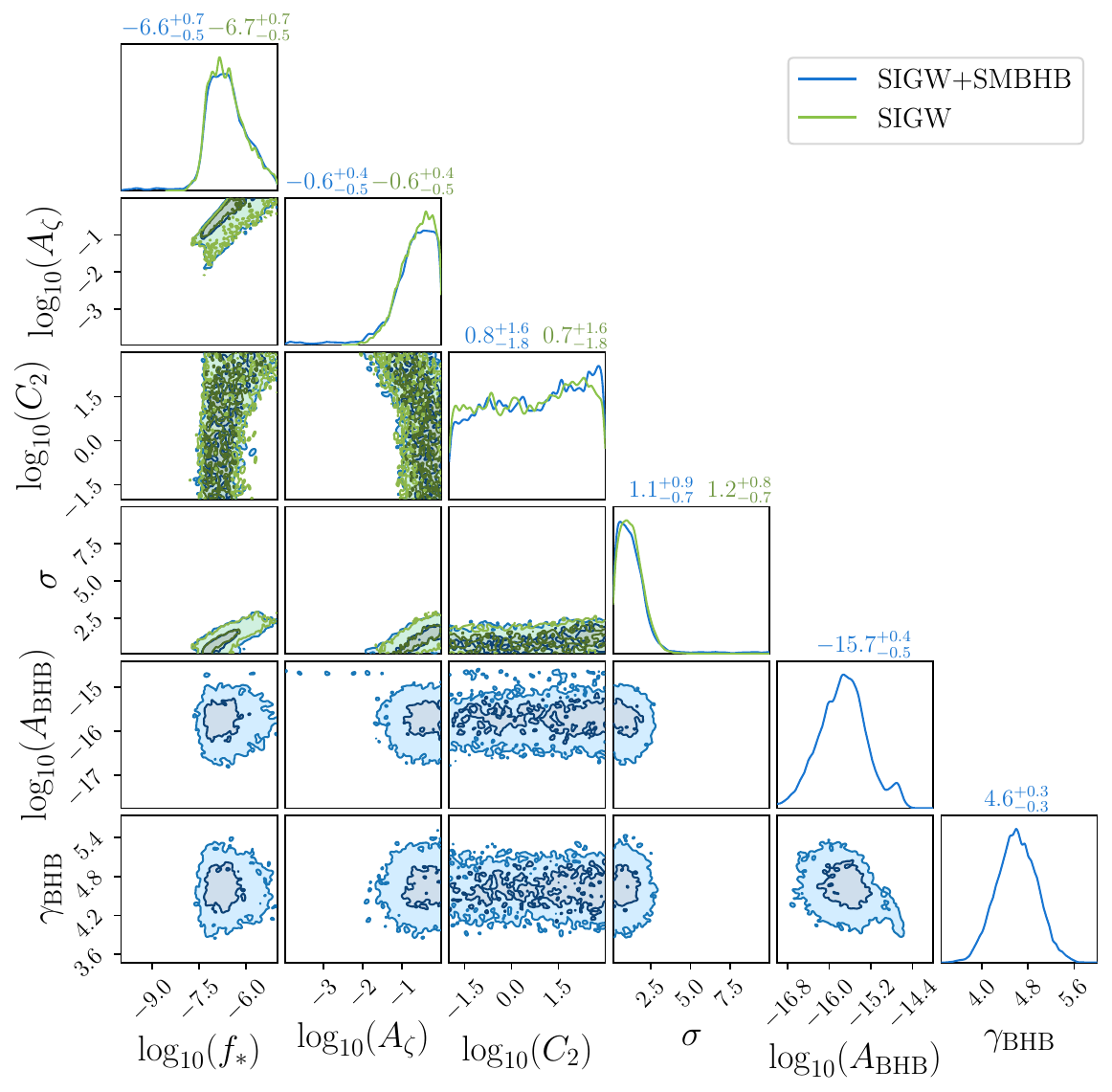}
\caption{Corner plot showing the posterior distributions with $C_2 \neq 0$ . The blue and green curves correspond to the NANOGrav 15-year dataset. Off-diagonal panels display $1$-$\sigma$ and $2$-$\sigma$ confidence intervals for joint distributions, and diagonal panels provide marginal distributions with the median values and $1$-$\sigma$ uncertainties noted above each histogram.} \label{fig:corner_cl2}
\end{figure}

\begin{figure}[htbp]
\includegraphics[width=0.5\linewidth]{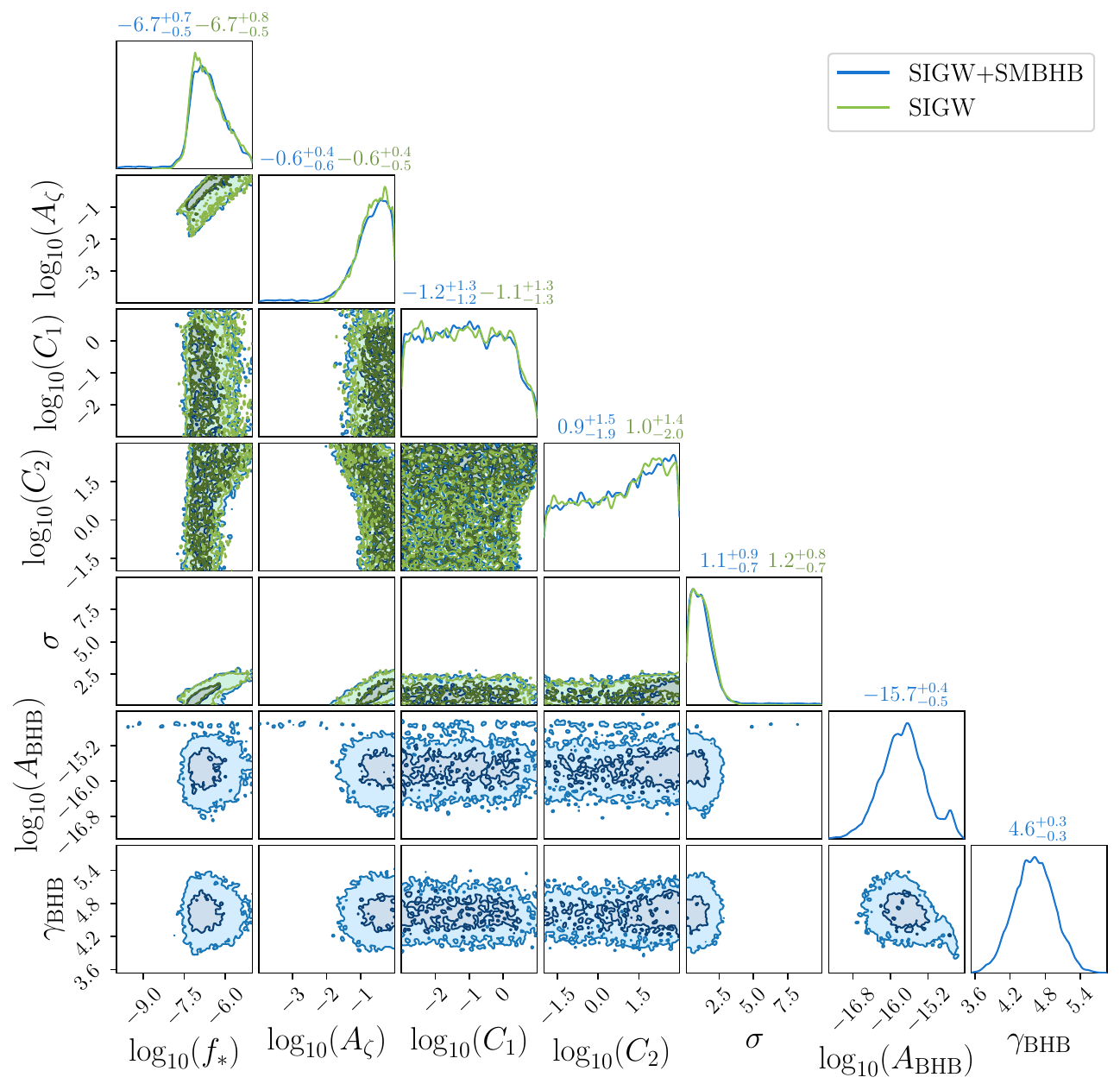}
\caption{Corner plot showing the posterior distributions with $C_{l} \neq 0$ $(l=1,2)$. The blue and green curves correspond to the NANOGrav 15-year dataset. Off-diagonal panels display $1$-$\sigma$ and $2$-$\sigma$ confidence intervals for joint distributions. Diagonal panels provide marginal distributions with the median values and $1$-$\sigma$ uncertainties noted above each histogram.} \label{fig:corner_cl1cl2}
\end{figure}

\begin{figure}[htbp]
\includegraphics[width=0.7\linewidth]{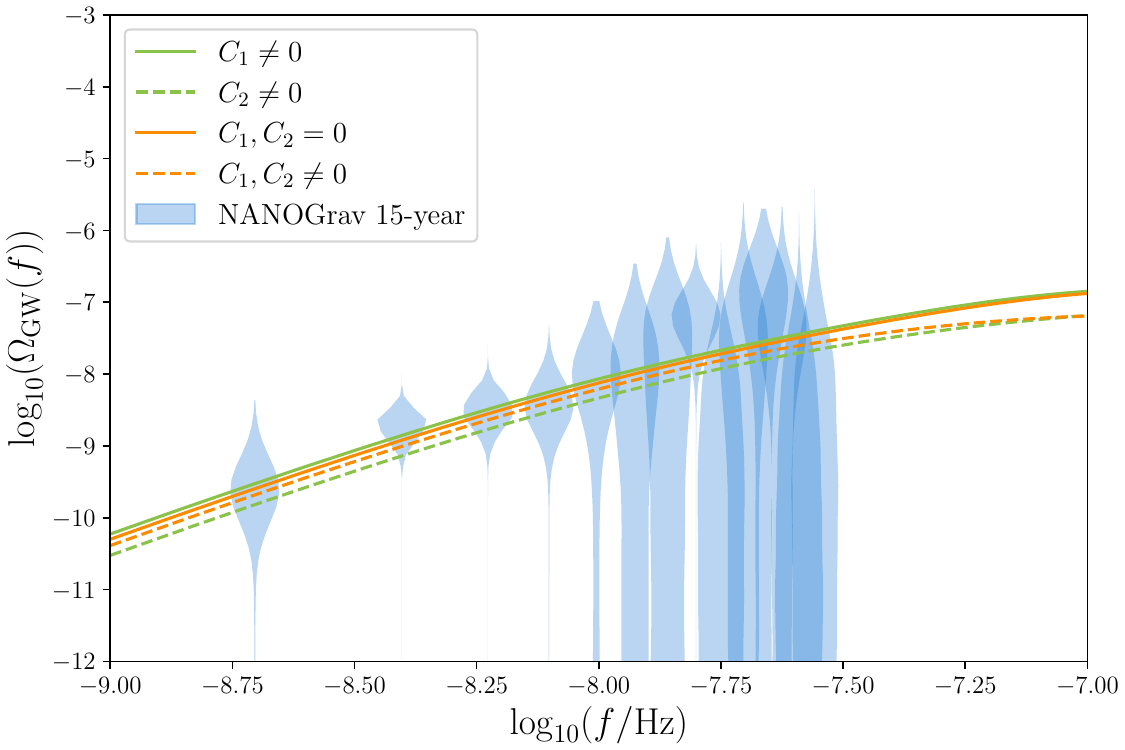}
\caption{The energy density spectra of second-order \acp{SIGW} with anisotropic primordial power spectra. The curves are based on parameters derived from the median values of the posterior distributions of the NANOGrav 15-year dataset.
The energy density spectra derived from the free spectrum of the NANOGrav 15-year dataset is shown with blue shading. Different line styles and colors indicate scenarios with $C_1, C_2 = 0$, $C_1 \neq 0$, $C_2 \neq 0$, and $C_1, C_2 \neq 0$, as labeled in the figure.} \label{fig:violinplot}
\end{figure}

To constrain the parameter space of the anisotropic primordial power spectrum using PTA observations, we employ the \ac{KDE} representations of the free spectra and construct the likelihood function~\cite{Lamb:2023jls,Moore:2021ibq,Mitridate:2023oar}:
\begin{equation} \label{eq:likelihood}
    \ln \mathcal{L}(d|\theta) = \sum_{i=1}^{N_f} p(\Phi_i,\theta)\ .
\end{equation}
In Eq.~(\ref{eq:likelihood}), $p(\Phi_i,\theta)$ represents the probability density of $\Phi_i$ for a given $\theta$, and $\Phi_i = \Phi(f_i)$ denotes the time delay:
\begin{equation} \label{eq:timedelay}
    \Phi(f) = \sqrt{\frac{H_0^2 \Omega_{\mathrm{GW}}(f)}{8\pi^2 f^5 T_{\mathrm{obs}}}} \ ,
\end{equation}
where $H_0=h\times 100 \mathrm{km/s/Mpc}$ is the present-day value of the Hubble constant.
Specifically, we directly utilize the publicly available KDE representations of the free spectrum provided by the NANOGrav 15-year dataset~\cite{Nanograv:KDE}. These KDEs are constructed based on an analysis explicitly accounting for Hellings--Downs (HD) spatial correlations. By adopting these official HD-correlated KDEs, our analysis strictly preserves the spatial information inherent in the PTA data, ensuring that the constraints are derived from a signal consistent with a \ac{GW} background rather than from uncorrelated common red noise.
Bayesian analysis is performed using \textsc{bilby}~\cite{bilby_paper} with its integrated \textsc{dynesty} nested sampler~\cite{Speagle:2019ivv,dynesty_software}. For the anisotropic primordial power spectra, the posterior distributions are presented in Fig.~\ref{fig:corner_cl1}--Fig.~\ref{fig:corner_cl1cl2}, with prior distributions for $\log_{10}(f_*/\mathrm{Hz})$, $\log_{10}(A_{\zeta})$, $\sigma$, $\log_{10}(C_1)$ and $\log_{10}(C_2)$ set as uniform over the intervals  $[-10, -5]$, $[-4, 0]$, $[0.1, 10]$, $[-3, 1]$, and $[-2, 3]$, respectively; here, we set $C_l=0$ $(l>2)$. The corresponding energy density spectra of second-order \acp{SIGW} are shown in Fig.~\ref{fig:violinplot}. As illustrated in Fig.~\ref{fig:corner_cl1}--Fig.~\ref{fig:corner_cl1cl2}, \ac{PTA} observations can effectively constrain the parameters $A_{\zeta}$, $\sigma$, and $f_*$, which that characterize the isotropic primordial power spectrum. However, owing to the presence of parameter degeneracies, the current PTA data cannot effectively constrain the parameters $C_1$ and $C_2$, which describe small-scale anisotropies. This result is mainly attributed to two factors. First, the current PTA observational data are not sufficiently precise. Second, the introduction of the anisotropy parameter $C_l$ leads to increased degeneracy in the energy density spectrum of \acp{SIGW}. To better constrain the primordial power spectrum anisotropy on small scales, additional cosmological observations that can place limits on the energy density spectrum of \acp{SIGW} must be considered.

\begin{figure}[htbp]

    \centering
    \includegraphics[width=0.5\columnwidth]{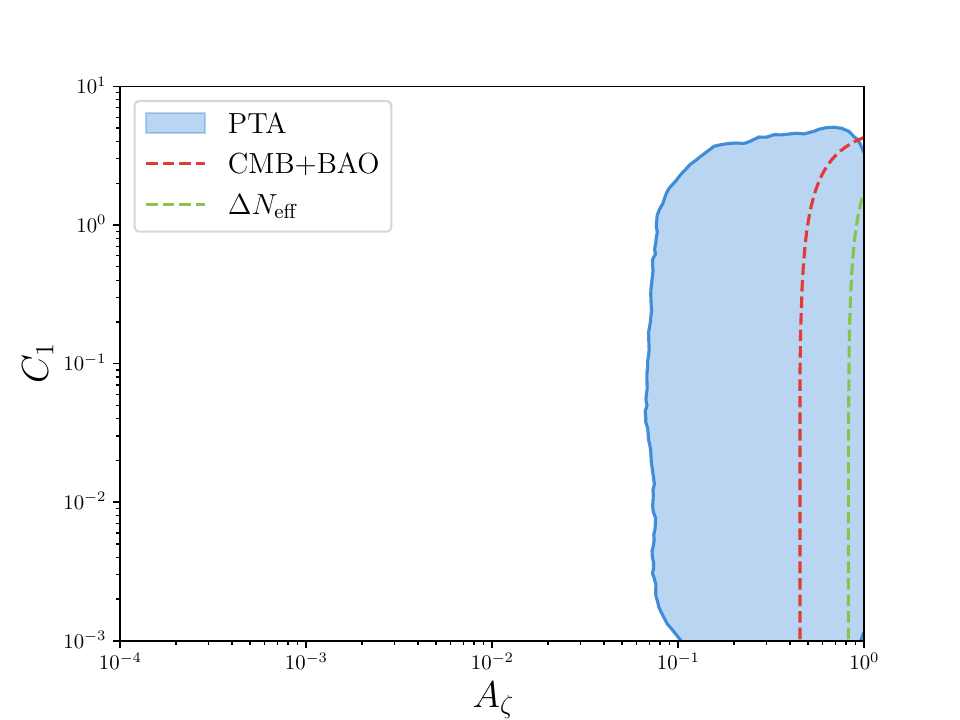}

\caption{\label{fig:constrain_AC1} Constraints on $A_\zeta$ and $C_1$ assuming $\sigma=1$ and $C_2=0$. The blue shaded region represents the $95\%$ credible intervals corresponding to the two-dimensional posterior distribution shown in Fig.~\ref{fig:corner_cl1} with the \ac{KDE} method. The red and green lines denote the lower bounds of $C_1$ from \ac{CMB} and \ac{BAO} observations (Eq.~(\ref{eq:rhup2})) and those from $\Delta N_{\mathrm{eff}}$ (Eq.~(\ref{eq:rhup1})), respectively.}
\end{figure}

\begin{figure}[htbp]

    \centering
    \includegraphics[width=0.5\columnwidth]{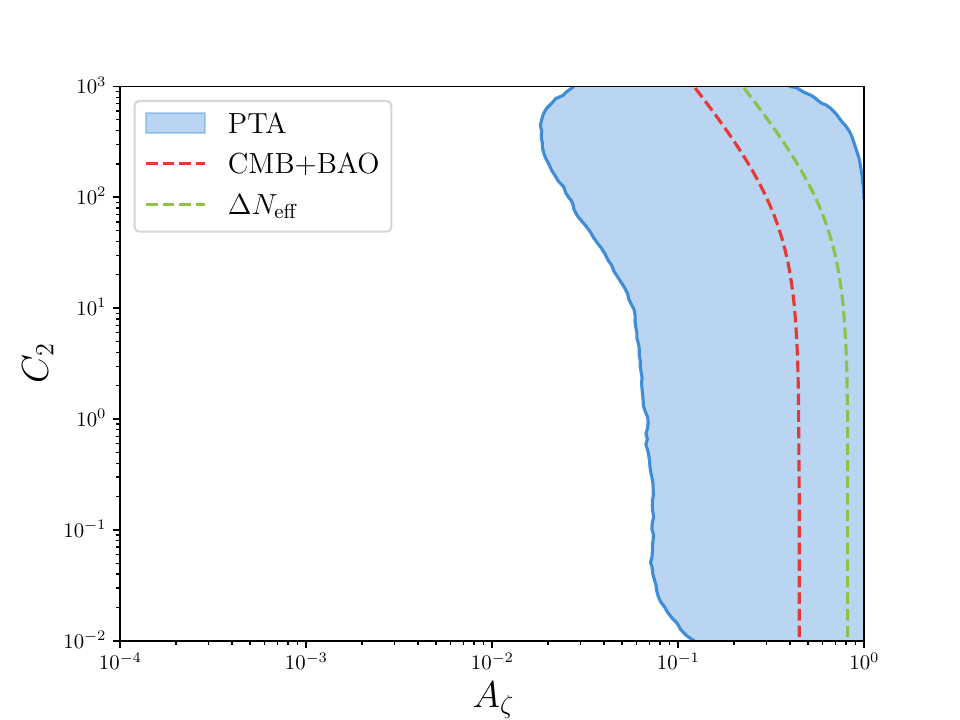}

\caption{\label{fig:constrain_AC2} Constraints on $A_\zeta$ and $C_2$ assuming $\sigma=1$ and $C_1=0$. The blue shaded region represents the $95\%$ credible intervals corresponding to the two-dimensional posterior distribution shown in Fig.~\ref{fig:corner_cl2} with the \ac{KDE} method. The red and green lines denote the lower bounds of $C_1$ from \ac{CMB} and \ac{BAO} observations (Eq.~(\ref{eq:rhup2})) and those from $\Delta N_{\mathrm{eff}}$ (Eq.~(\ref{eq:rhup1})), respectively.}
\end{figure}

\begin{figure}[htbp]

    \centering
    \includegraphics[width=0.5\columnwidth]{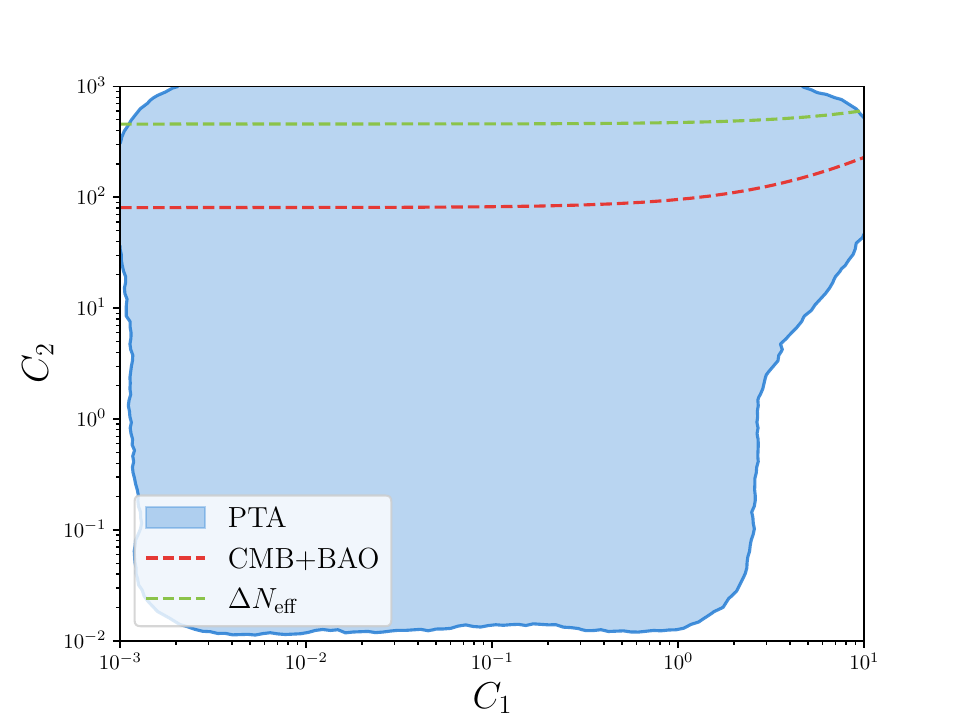}

\caption{\label{fig:constrain_C12} Constraints on $C_1$ and $C_2$ assuming $\sigma=1$ and $A_\zeta=10^{-0.5}$. The blue shaded regions represent the $95\%$ credible intervals corresponding to the two-dimensional posterior distribution shown in Fig.~\ref{fig:corner_cl1cl2} with the \ac{KDE} method. The black solid line and gray dashed line denote the upper bounds of $C_2$ from \ac{CMB} and \ac{BAO} observations (Eq.~(\ref{eq:rhup2})) and those from $\Delta N_{\mathrm{eff}}$ (Eq.~(\ref{eq:rhup1})), respectively.}
\end{figure}

In this section, we consider the constraints from large-scale cosmological observations on second-order \acp{SIGW} and the corresponding parameter space of the small-scale anisotropic primordial power spectrum. Large-scale cosmological observations constrain the \ac{SGWB} via two types of methods. The first method treats the \ac{SGWB} as an additional radiation component, modifying the effective number of relativistic species, $N_{\mathrm{eff}}$. To remain consistent with current observational bounds, the total energy density spectrum must satisfy~\cite{Ben-Dayan:2019gll,Cang:2022jyc}
\begin{eqnarray}\label{eq:rhup1}
  \int_{f_{\mathrm{min}}}^{\infty} h^2\Omega_{\mathrm{GW},0}(k) \mathrm{d} \left(\ln k\right) < 1.3 \times 10^{-6}\frac{\Delta N_{\mathrm{eff}}}{0.234} \   ,
\end{eqnarray}
 where $\Delta N_{\mathrm{eff}}= N_{\mathrm{eff}}-3.046$. Here, we use the $N_{\mathrm{eff}}$ limits provided by Aghanim \emph{et al}.~\cite{Planck:2018vyg}, who report $N_{\mathrm{eff}}=3.04 \pm 0.22$ at a $95\%$ confidence level for the \texttt{ Planck} + \ac{BAO} + \ac{BBN} data. The second method directly utilizes \ac{CMB} and \ac{BAO} measurements, imposing the following stricter constraint:
 \begin{equation}\label{eq:rhup2}
\int_{f_{\mathrm{min}}}^{\infty} h^2\Omega_{\mathrm{GW},0}(k) \mathrm{d}\left(\ln k\right) < 2.9\times 10^{-7}  \ ,
\end{equation}
at $95\%$ confidence level for \ac{CMB}$+$\ac{BAO} data~\cite{Clarke:2020bil}. Fig.~\ref{fig:constrain_AC1}--Fig.~\ref{fig:constrain_C12} present the current constraints from large-scale cosmological observations on the parameter space composed of $C_1$ and $C_2$. The green and the red lines correspond to the observational limits given by Eq.~(\ref{eq:rhup1}) and Eq.~(\ref{eq:rhup2}), respectively.

\begin{figure}[htbp]
    \centering
    \includegraphics[width=0.8\columnwidth]{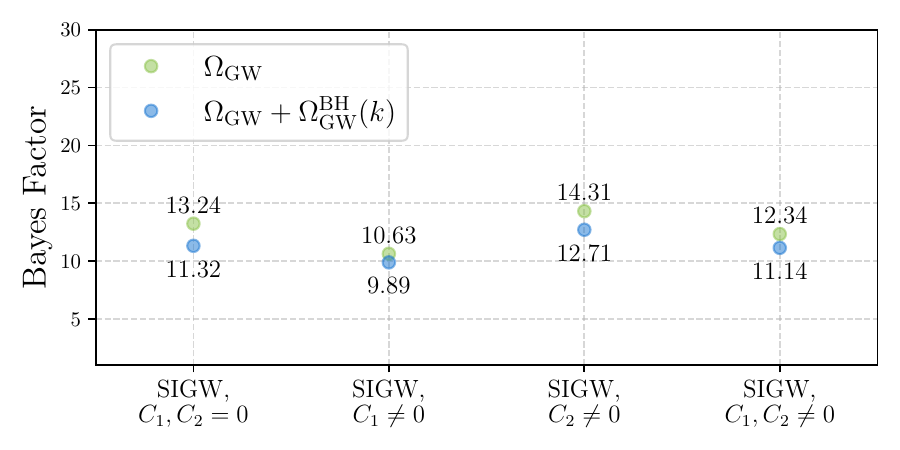}
\caption{\label{fig:bayes} Bayes factors between different models. The vertical axis represents the Bayes factor of different models relative to \ac{SMBHB}, and the horizontal axis represents the different models. The green dots corresponding to models without \ac{SMBHB}, and the blue dots represent models in combination with the \ac{SMBHB} signal. }
\end{figure}

Furthermore, to better assess the plausibility of different models in explaining current \ac{PTA} observations, we perform a detailed analysis of the Bayes factors between different models. The Bayes factor is defined as $B_{i,j} = \frac{Z_i}{Z_j}$, where $Z_i$ represents the evidence of model $H_i$. In addition to \acp{SIGW}, we investigate a hybrid model scenario, wherein both \acp{SMBHB} and \acp{SIGW} contribute jointly. The energy density spectrum of SMBHBs is characterized by~\cite{NANOGrav:2023hvm}
\begin{equation} \label{eq:SMBHB}
    \Omega_{\mathrm{GW}}^{\mathrm{BH}}(f) = \frac{2\pi^2 A_{\mathrm{BHB}}^2}{3H_0^2 h^2} (\frac{f}{\mathrm{year}^{-1}})^{5-\gamma_{\mathrm{BHB}}}\mathrm{year}^{-2} \ ,
\end{equation}
with the prior distribution for $\log_{10}A_{\mathrm{BHB}}$ assumed to follow a multivariate Gaussian distribution~\cite{NANOGrav:2023hvm}
\begin{equation} \label{eq:prior_SMBHB}
\begin{aligned}
    \boldsymbol{\mu}_{\mathrm{BHB}}&=\begin{pmatrix} -15.6
 \\ 4.7 \end{pmatrix} \ , \\
\boldsymbol{\sigma}_{\mathrm{BHB}}&=0.1\times \begin{pmatrix}
2.8  & -0.026\\
-0.026  & 2.8
\end{pmatrix} \ .
\end{aligned}
\end{equation}
As shown in Fig.~\ref{fig:bayes}, we calculate the Bayes factors between different models. The results indicate that when small-scale anisotropies in the primordial power spectrum are considered, the Bayesian factor for \acp{SIGW} within the considered parameter space is approximately $10$. This result suggests that compared to the \ac{SMBHB} model, \acp{SIGW} with different anisotropic parameters are more likely to dominate current \ac{PTA} observations. To directly assess the statistical preference for small-scale anisotropy, the Bayes factor between the anisotropic models and isotropic \ac{SIGW} model ($C_1=C_2=0$) can be derived simply by taking the ratio of the Bayes factors presented in Fig.~\ref{fig:bayes}. The Bayes factor for the most favored anisotropic case ($C_2 \neq 0$) relative to the isotropic case is approximately $\mathcal{B} \approx 14.31/13.24 \approx 1.08$, which is close to unity. This value indicates that current PTA data are insufficient to effectively distinguish between isotropic and anisotropic primordial power spectra on small scales.

{
\subsection{Constraints from future observations and specific models}\label{sec:4.2}

 In Sec.~\ref{sec:4.1}, we investigate the impact of second-order \acp{SIGW}, generated by small-scale anisotropic primordial power spectra, on current PTA observations. By combining PTA data with large-scale cosmological observations, we constrain the small-scale anisotropy parameters. Our results show that when considering a general parameterized form of the anisotropic primordial power spectrum, the presence of anisotropy parameters leads to a certain degree of degeneracy in the energy density spectrum of \acp{SIGW}. Consequently, current cosmological observations cannot impose stringent constraints on small-scale anisotropy parameters. To address this issue, we jointly constrain or determine the parameter space of the small-scale primordial power spectrum by combining \ac{SGWB} observations at different frequency bands.

 As shown by the posterior distributions in Fig.~\ref{fig:corner_cl1cl2}, when second-order \acp{SIGW} dominate current PTA observations, the parameters $A_{\zeta}$, $\sigma$, and $f_*$ of the isotropic primordial power spectrum $\mathcal{P}_{0,\zeta}$ can be constrained relatively effectively. However, the anisotropy parameters $C_1$ and $C_2$ remain unconstrained. In this analysis, we fix the isotropic parameters of $\mathcal{P}_{0,\zeta}$ to the median values determined by PTA observations (Fig.~\ref{fig:corner_cl1cl2}) and vary the anisotropy parameters $C_1$ and $C_2$ to examine their impact on \ac{SGWB} observations in the \ac{LISA} frequency band. More precisely, to asssess the impact of the anisotropic primordial power spectrum on second-order \acp{SIGW}, we calculate the \ac{SNR} $\rho$ for \ac{LISA}~\cite{Siemens:2013zla, Robson:2018ifk}:
\begin{equation}
    \rho = \sqrt{T}\left[ \int \mathrm{d} f\left(\frac{\bar{\Omega}_{\mathrm{GW},0}(f)}{\Omega_\mathrm{n}(f)}\right)^2\right]^{1/2} \ ,
\end{equation}
where $T$ is the observation time, and we set $T=4$ years here. $\Omega_\mathrm{n}(f)=2\pi^2f^3S_n/3H_0^2$, where $H_0$ is the Hubble constant, and $S_n$ is the strain noise power spectral density~\cite{Robson:2018ifk}.
\begin{figure}[htbp]
\includegraphics[width=1\linewidth]{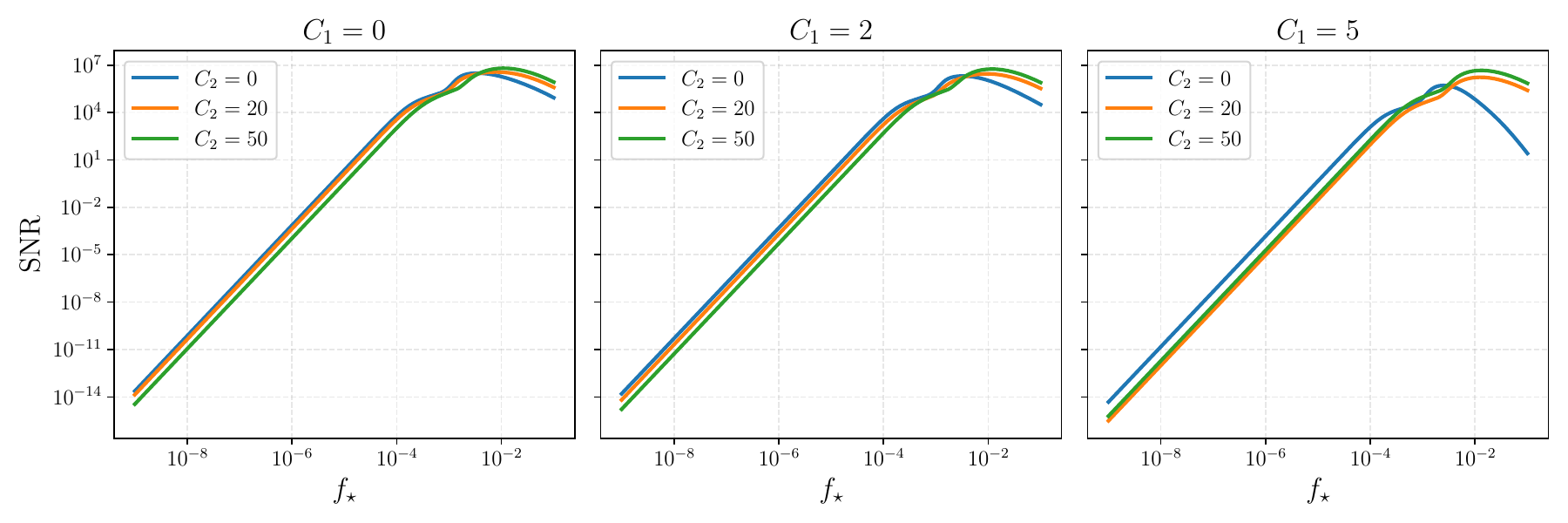}
\caption{The \ac{SNR} of \ac{LISA} as a function of $f_*$ with different $C_1$ and $C_2$, calculated for fixed values of $A_\zeta = 10^{-0.6}$ and $\sigma = 1$. The gray bands represent the $1$-$\sigma$ range from the posterior of $f_*$ shown in Fig.~\ref{fig:corner_cl1cl2}. \label{fig:snr_cl1cl2}}
\end{figure}
As illustrated in Fig.~\ref{fig:corner_cl1cl2}, if the second-order \acp{SIGW} generated by an anisotropic primordial power spectrum dominate the current PTA observations, then $f_*\approx 10^{-6.7}$ Hz, $A_{\zeta}\approx 10^{-0.6}$, and $\sigma\approx 1.2$. In Fig.~\ref{fig:snr_cl1cl2}, we present the \acp{SNR} of \ac{LISA} corresponding to different anisotropy parameters $C_1$ and $C_2$ under this scenario. As shown in Fig.~\ref{fig:snr_cl1cl2}, the \ac{SNR} of \ac{LISA} increases with the peak position $f_*$ of the primordial power spectrum. When $f_*>10^{-3}$, the anisotropy parameters $C_1$ and $C_2$ significantly influence the \ac{SNR} of \ac{LISA}. Thus, future observations of the \ac{SGWB} in the \ac{LISA} frequency band may provide a new window for studying anisotropic inflation models at small scales.

Furthermore, the gray region in Fig.~\ref{fig:corner_cl1cl2} corresponds to the $1$-$\sigma$ range of the posterior of $f_*$ when \acp{SIGW} dominate current \ac{PTA} observations. When $f_*$ lies within this gray region, irrespective of the values of $C_1$ and $C_2$, the \ac{SNR} of \ac{LISA} does not exceed $10^{-2}$. Therefore, the \acp{SIGW} generated by the primordial power spectrum under consideration cannot simultaneously dominate \ac{PTA} observations and significantly influence the \ac{SGWB} in the \ac{LISA} band. This behavior is markedly different from that of the \ac{SGWB} produced by \ac{SMBHB}. When the \ac{SMBHB} model dominates the \ac{PTA} signal, the corresponding energy density spectrum can still be detected by \ac{LISA}~\cite{Ellis:2023oxs}. Combining \ac{SGWB} observations across different frequency bands to jointly constrain or determine various \ac{SGWB} models and their corresponding parameter spaces may serve as an important tool in the future for distinguishing between different sources of the \ac{SGWB}.

The aforementioned discussion reveals that current cosmological observations are insufficient to impose strong constraints on the small-scale anisotropic primordial power spectrum. More precise future observations, combined with \ac{SGWB} measurements at other frequency bands, are required to better constrain or determine the properties of the small-scale anisotropic primordial power spectrum. Notably, the small-scale anisotropic primordial power spectra considered here are given in a parameterized form, in which the anisotropy parameter $C_l$ is not subject to any restriction. However, in specific inflationary models, the anisotropy parameters are generally dictated by the intrinsic features of the model; thus the parameter $C_l$ is not completely free. As an illustration, Yokoyama and Soda~\cite{Yokoyama:2008xw} considered a system constructed by two scalar fields, specifically, an inflaton $\phi$ and a waterfall field $\chi$, and a vector field $A_\mu(\mu=0,1,2,3)$, which coupled with the waterfall field. The corresponding action can be written as
\begin{equation}
\begin{aligned}
S=\frac{1}{2} &\int \mathrm{d}^4 x \sqrt{-g} R-\int \mathrm{d}^4 x \sqrt{-g}\left[\frac{1}{2} g^{\mu \nu}\left(\partial_\mu \phi \partial_\nu \phi+\partial_\mu \chi \partial_\nu \chi\right)+V\left(\phi, \chi, A_\mu\right)\right] \\
&-\frac{1}{4} \int \mathrm{d}^4 x \sqrt{-g} g^{\mu \nu} g^{\rho \sigma} f^2(\phi) F_{\mu \rho} F_{\nu \sigma} \ ,
\end{aligned}
\end{equation}
where $F_{\mu \nu} \equiv \partial_\mu A_\nu-\partial_\nu A_\mu$ is the field strength of the vector field, and an arbitrary function $f(\phi)$ represents gauge coupling. The potential of field $V\left(\phi, \chi, A_\mu\right)$ is given by
\begin{equation}
V\left(\phi, \chi, A^i\right)=\frac{\lambda}{4}\left(\chi^2-v^2\right)^2+\frac{1}{2} g^2 \phi^2 \chi^2+\frac{1}{2} m^2 \phi^2+\frac{1}{2} h^2 A^\mu A_\mu \chi^2 \ .
\end{equation}
In such a scenario, the anisotropic primordial power spectrum can be represented as
\begin{equation}
    \begin{aligned}   \mathcal{P}^{\hat{\mathbf{n}}}_\zeta(\mathbf{k}) = \mathcal{P}_{0,\zeta}(k)\left(1-g_*\left(\hat{\mathbf{n}}\cdot\hat{\mathbf{k}}\right)^2\right) \  ,
    \end{aligned}
\end{equation}
where
\begin{equation}
g_*=\frac{\beta}{1+\beta} \ , \ \beta \approx \frac{1}{f_e^2}\left(\frac{h^2|\mathbf{A}|}{g^2 \phi_e}\right)^2 \ .
\end{equation}
In this model, $g_*$ serves as the parameter quantifying anisotropy. When $\beta\gg 1$, $g_*\approx 1$, whereas when $\beta\ll 1$, $g_*\approx\beta\ll 1$. In this case, irrespective of how the model parameters vary, generating an extremely large anisotropic primordial power spectrum is impossible. The above example illustrates that in addition to the constraints from cosmological observations, specific anisotropic inflationary models may impose further restrictions on the anisotropy parameters. For a given model, current cosmological data can be combined with the properties of the model to jointly constrain the parameter space of the small-scale anisotropic primordial power spectrum.
}

\section{Conclusion}\label{sec:5.0}
In this study, we explored the effect of small-scale anisotropic primordial power spectra on the energy density spectrum of second-order \acp{SIGW}. In Sec.~\ref{sec:3.0}, we derive the expression for the energy density spectrum of these waves under the influence of an anisotropic primordial power spectrum. This result is applicable to any form of small-scale anisotropic primordial power spectrum with $l \leq 4$ and remains independent of specific inflation models. As examples, we examine the gauge field and Finslerian inflation models, which correspond to the primordial power spectra with $C_2\neq 0$ and $C_1\neq 0$, respectively. We investigate the impact of these two inflation models on current \ac{PTA} observations and \ac{SNR} of \ac{LISA}, and extend the findings to a general scenario, wherein both $C_1$ and $C_2$ are non-zero.

In Sec.~\ref{sec:4.0}, we discuss the constraints imposed by current \ac{PTA}$+$\ac{CMB}$+$\ac{BAO} data on the small-scale anisotropic primordial power spectrum. Specifically, under the assumption that second-order \acp{SIGW} dominate current PTA observations, we analyze the constraints imposed by PTA data on small-scale anisotropic inflation models. To better constrain the parameter space of different models, we incorporate large-scale cosmological observations and present the resulting constraints on the parameter space of the small-scale anisotropic primordial power spectrum in Fig.~\ref{fig:constrain_AC1}--Fig.~\ref{fig:constrain_C12}. In addition, we examine the feasibility of \acp{SIGW}, generated by different small-scale anisotropic inflation models, dominating current PTA signals. The results indicate that current cosmological data can only partially constrain the parameter space of these models, without definitively confirming or ruling them out.

Because current cosmological observations cannot effectively constrain the general form of the anisotropic primordial power spectrum, we further consider the restrictions from future \ac{SGWB} observations in the LISA frequency band, as well as from specific inflationary models. Using the isotropic primordial power spectrum parameters determined by PTA observations, we analyze the impact of anisotropy parameters $C_1$ and $C_2$ on the \ac{SNR} of \ac{LISA} under the scenario, wherein \acp{SIGW} dominate current PTA observations. The results show that the \acp{SIGW} generated by the primordial power spectrum under consideration cannot simultaneously dominate \ac{PTA} observations and significantly impact the \ac{SGWB} in the \ac{LISA} band. Furthermore, we briefly analyze anisotropic primordial power spectra generated by specific models, noting that in certain anisotropic inflationary scenarios, the models themselves impose relatively strict constraints on the anisotropy parameters.

\vspace{0.3cm}
\begin{acknowledgements}
The work is supported in part by the National Natural Science Foundation of China (NSFC) grants No.12475075, No.12447127
\end{acknowledgements}

\bibliography{biblio}

\end{document}